\documentclass[referee,useAMS,usenatbib,usegraphicx]{mn2e}
\usepackage{amssymb}
\usepackage{graphicx}
\usepackage{rotating}
\usepackage{lscape}

\title[Bok globule CB17: Polarization, extinction and distance]
{Bok globule CB17: Polarization, extinction and distance}
\author[G. B. Choudhury, A. Barman,  H. S. Das and  B. J. Medhi ]
{G. B. Choudhury$^{1}$\thanks{E-mail: gulafsha.97@gmail.com (GBC)}
, A. Barman$^{1}$\thanks{Email: ajoy2bar@gmail.com (AB)}
, H. S. Das$^{1}$\thanks{E-mail: hsdas13@gmail.com (HSD)}
and B. J. Medhi$^{2}$\thanks{E-mail: biman@gauhati.ac.in (BJM)}
\\
$^{1}$ Department of Physics, Assam University, Silchar 788011, India.\\
$^{2}$ Department of Physics, Gauhati University, Guwahati 781014, India\\
}

\begin{document}

\pagerange{\pageref{firstpage}--\pageref{lastpage}} \pubyear{2018}

\maketitle

\label{firstpage}

\begin{abstract}
In this paper, the results obtained from the polarimetric study of a Bok globule CB17 in both optical and sub-millimeter wavelength are presented. The optical polarimetric observations in R-band ($\lambda$ = 630 nm, $\Delta \lambda$ = 120 nm) were conducted from 1.04-meter Sampurnanand Telescope, ARIES, Nainital, India on 9th March 2016, while, the sub-mm polarimetric data are taken from the SCUPOL data archive which has been reanalyzed. The contours of Herschel\footnote{Herschel is an ESA space observatory with science instruments provided by European-led Principal Investigator consortia and with important participation from NASA.} SPIRE 500$\mu m$ dust continuum emissions of CB17 (typically a cometary-shaped globule) are overlaid on the DSS image of CB17 along with polarization vectors (optical and submm). The magnetic field strength at the core of the globule is estimated to be $99\mu$G. Using the Near-Infrared photometric technique and \emph{Gaia} data, the distance to CB17 is found to be $253 \pm 43$ parsec. A correlation between the various quantities of the globule is also studied. It is observed that the magnetic field in the cloud core as revealed by polarization measurements at the sub-millimeter dust emission is found to be almost aligned along the minor axis of the globule which fits the magnetically regulated star formation model. The misalignment between core-scale magnetic field direction and molecular outflow direction is also found.
\end{abstract}


\section{Introduction}

The formation of stars in our galaxy is a result of collapse and fragmentation in giant molecular clouds. But, due to the complexity, turbulence and multiple episodes of star formation, these are not considered as the best objects to study the early stages of star formation (Yun and Clemens 1991). Bok globules, being the simplest, isolated molecular clouds (dark clouds) in our Milky Way galaxy are excellent sites for low mass star formation (Bok and Reilly 1947). Bok globules were first observed by astronomer Bart J Bock in 1940s; these are small, opaque and relatively isolated molecular clouds with diameters of about 0.7 pc (0.1--2 pc), the temperature $10-15$ K, the density $10^4-10^5$ cm$^{-3}$ and masses of $\approx10M_\odot (2-100M_\odot)$ (Bok 1977 and Leung 1985). Polarimetric observation of these clouds at optical wavelength can give the information about the magnetic field orientation in the low-density edge region of clouds. However, the polarimetric observations in IR and sub-millimeter wavelength can range the magnetic field orientation in high density central region of clouds (Kane et al. 1995; Alves et al. 2008; Ward-Thompson et al. 2009; Franco et al. 2010 ; Paul et al. 2012; Chakraborty et al. 2014; Bertrang et al. 2014; Chakraborty \& Das 2016; Das et al. 2016; Soam et al. 2016 etc.).

The study of magnetic field is important as it plays a major role in the evolution of dark clouds and may control the fragmentation of clouds to form stars (Mestel and Spitzer 1956; Nakano and Nakamura 1978; Mouschovias \& Morton 1991; Li \& Nakamura 2004). Previously our group performed the polarimetric study of two Bok Globules CB 34 (Das et al. 2016) and CB 130 (Chakraborty and Das 2016) which enriched the knowledge about the magnetic field geometry of two globules. Das et al. (2016) studied magnetic field regions near two submm cores C$_1$ and C$_2$, and estimated the magnetic field strength of two cores. They also reported a correlation between the mean magnetic field, the minor axis and outflow direction for both the cores of the cloud. Chakraborty and Das (2016) studied the orientation of the local magnetic field of the cloud CB130 and reported an offset of 53$^\circ$ between the orientation of the envelope magnetic field and the galactic magnetic field of the cloud. To understand the star formation process well, the detailed properties of cores and their surroundings at different evolutionary stages have to be studied (Chen et. al. 2012). Observations from the \emph{Herschel} Space Observatory (Pilbratt et al. 2010) at 100 -- 500$\mu m$ had provided the excellent resolution and wavelength coverage to accurately map the density structure of different globules from thermal dust emission. Recently, Il'in et al. (2018) studied the outer layers and vicinity of the B5 globule using \emph{UBVRI} polarimetric observations and data available from different sources and surveys. They used the Gaia\footnote{\textit{Gaia} is a ESA mission, launched on 19 December 2013, which provided exceptional measurements of the positions, motions, and distances of more than one billion stars in the Milky Way Galaxy.} parallaxes data of stars in direction towards B5 together with extinction measurements to constrain a 3D extinction map.

In this paper, we present the optical and sub-millimeter polarimetric analysis of the globule CB 17 to study the magnetic field in both low as well as in high density region of the globule, thereby estimating the strength of magnetic field at the core of the cloud. We also have made use of $Gaia$ data to revisit the distance to this cloud using the Near-Infrared photometric technique. We have also studied the relative orientations between various quantities of CB17 and made a comparative study of the same for some other dark clouds.

\section{Target Globule: CB 17}
CB17 (other name L1389) is a cometary shaped small globule, which has a dense submm core (SMM) and a faint cold IRAS point source (IRAS 04005+5647, detected only at 60 $\mu m$ and 100 $\mu m$) (Launhardt et al. 2010). The  deep 1.3 mm continuum map suggests that there may be a double cores (SMM1 and SMM2) in CB17 SMM with 14$''$ separation and a common envelope located at the southwestern edge of the cloud head, although this result was not detected in SCUBA 850 $\mu m$ images (see Launhardt et al. 2010). Das et al. (2015) estimated the distance of CB17 using near-infrared photometry which is given by 478 $\pm$ 88 parsec.  It is located near Perseus and is associated with Lindblad ring (Lindblad et al. 1973). A cometary-shaped morphology of the cloud has been observed in Herschel-SPIRE (Spectral and Photometric Imaging Receiver; Griffin et al. 2010) images of CB17 at 500 $\mu m$ (Launhardt et al. 2013). Optical polarimetric observations of CB17 were also conducted by Soam et al. (2016) to understand the envelope magnetic field. This globule was extensively studied by other investigators which enriched the knowledge about this globule (Kane \& Clemens 1997; Benson et al. 1998; Pavlyuchenkov et al. 2006; Chen et al. 2012; Schmalzl et al. 2014 etc.). The basic parameters related to the source is presented in Table-1.

Polarimetric observations of a globule at sub-millimeter wavelength can trace the magnetic field orientation towards the high-density central region of the clouds. We have chosen the cloud CB17, particularly, because of the availability of the sub-millimeter data, which in addition with optical analysis will give a better exposure to understand the magnetic field geometry over that region of the sky. Study of magnetic field is important as it plays a major role in the star formation processes. As CB17 is located near the galactic plane, the magnetic field in this region of the galaxy will admit us to constraint how close to the Bok globule the environmental magnetic field is ruling over and where the magnetic field of the globule is preferably governed by its internal field.


\begin{table*}
\caption{Basic parameters related to the source studied. Right ascension (RA), Declination (DEC), galactic longitude and latitude ($l,b$), the mean angular size, the mean volume density of the core ($n_{H_2}$), position angle of galactic plane ($\theta_{GP}$), distance of the cloud, position angle of outflow axis ($\theta_{out}$) and position angle of the minor axis ($\theta_{min}$).}
\begin{center}
\begin{tabular}{|c|c|c|c|c|c|c|c|c|c|c|}
\hline
  RA(2000)$^{a}$ & DEC(2000)$^{a}$   		& ($l,b$)      & \multicolumn{2}{c}{Size$^{b}$} & $n_{H_2}$$^{c}$   & $\theta_{GP}$	& Distance		  & $\theta_{out}$$^{e}$ &  $\theta_{min}$$^{f}$  \\
	
	\cline{4-5}\\[-0.1cm]

	(h m s) 			& ($^\circ$ $'$ $''$)   & ($^\circ$)   & ($'$) & (pc)     & ($cm^{-3}$) & ($^\circ$)     & (parsec) &  ($^\circ$)		     &  ($^\circ$)      \\
 \hline

  04~04~38       &  56~56~12  & 147.02, 3.39 & 2.2  & 0.16       & 3.9$\times10^5$ & 132            & 478 $\pm$ 88$^{d}$  &  125   & 50       \\
								&									   &							&	        &   & &		             & 250 $\pm$ 50$^{b}$	  &					&          			\\

 \hline
\end{tabular}
\end{center}
$^{(a)}$Launhardt et al. (2013)
$^{(b)}$Launhardt et al. (2010)
$^{(c)}$Launhardt et al. (1997)
$^{(d)}$Das et al. (2015)
$^{(e)}$Chen et al. (2012)
$^{(f)}$Soam et al. (2016)
\end{table*}



\begin{table*}
\caption{Observation log of CB17.}
\begin{center}
\begin{tabular}{|c|c|c|c|c|c|}
\hline
   Object ID & Name of         &Date           & Fields & RA(2000) & DEC(2000)           \\
             & Observatory     &               &        & (h m s)  & ($^\circ$ $'$ $''$)  \\
 \hline
   CB17      & ARIES, Nainital &  March 9,2016 &  F1    & 04 04 31 & 56 56 49 					 \\
             & ARIES, Nainital &               &  F2    & 04 04 32 & 56 54 16  					 \\
					   & ARIES, Nainital &               &  F3    & 04 04 59 & 56 55 14 					 \\
 \hline
\end{tabular}
\end{center}
\end{table*}


\section{Observation}

The polarimetric observation of CB17 was conducted on 9th March, 2016 using 1.04-meter Sampurnanand f/13 Cassegrain Telescope, ARIES(Aryabhatta Research Institute of observational sciencES), Nainital, India which uses an imaging polarimeter AIMPOL (ARIES IMaging POLarimeter). The details are well presented in Rautela et al. (2004), Medhi et al. (2010) and Das et. al. (2016). Polarimetric observations of Bok Globule CB17 carried out at broadband R-filter ($\lambda$ = 630 nm, $\Delta \lambda$ = 120 nm) were made for three different sub-regions F1, F2 and F3 (the details of three fields of CB17 are shown in Table-2.) to cover CB17 core, particularly the South and West region of the cloud, as the Northern region has already been observed by Soam et al. 2016 (in optical R-band). Our objective is to study the magnetic field geometry in the low density region of the cloud surrounding CB17.

The instrumental calibration for both the polarization and zero position angle is done by observing an un-polarized standard star GD319 and two polarized standard stars HD 19820 and HD 25443 respectively.


\begin{table*}
\caption{Optical polarization results of 24 stars observed towards CB17. Column--2 and 3 give the RA and DEC of the stars, column-4 and 5 represent the magnitude and position angle of polarization obtained from our study. Column--6 gives whether the value of $p/ep$ is greater than 3 or not. Column--7 and 8 represent the magnitude and position angle of polarization which are collected from Soam et al. (2016). }
\begin{center}
\begin{tabular}{|c|c|c|c|c|c|c|c|}
\hline
   Star ID & RA(2000)($^\circ$) & Dec(2000)($^\circ$) & $p$ $\pm$ ep ($\%$) & $\theta$ ($^\circ$) & $p/ep>3$ & $p^\prime$$\pm ep^\prime$ ($\%$) &  $\theta^\prime$($^\circ$) $^\dag$ \\
 \hline
1  & 61.326 &	56.869	&	4.26$\pm$1.08		&		138.1	& Y	&	$-$	&	$-$	\\
2  & 61.298	&	56.867	&	0.92$\pm$0.24		&		139.9	& Y	&	$-$	&	$-$	\\
3  & 61.279	&	56.870		& 3.43$\pm$1.04		&		138.9	& Y	&	$-$	&	$-$	\\
4  & 61.236	&	56.862	&	3.93$\pm$1.23		&		136		& Y	& $-$	&	$-$	\\
5	 & 61.193	&	56.822	&	3.48$\pm$0.51		&		136.2 & Y		&	$-$	&	$-$	\\
6	 & 61.176	&	56.988	&	2.61$\pm$0.59		&		130.1	& Y	&	$-$	&	$-$	\\
7	 & 61.167	&	56.859	&	3.95$\pm$1.08	 	&		141.6	& Y	&	$-$	&	$-$	\\
8	 & 61.138	&	56.962	&	3.66$\pm$0.86		&		132.9	& Y	&	3.6$\pm$0.3	&	134	\\
9	 & 61.120 &	56.913	&	2.24$\pm$0.50		&		137.8	& Y	&	2.1$\pm$0.1	&	136	\\
10 & 61.115	&	56.941	&	4.72$\pm$1.33		&		139.7	& Y	&	4.2$\pm$0.2	&	139 \\
11 & 61.103	&	56.847	&	2.99$\pm$0.44		&		135.3	& Y	&	$-$	&	$-$	\\
12 & 61.076	&	57.006	&	3.36$\pm$1.05		&		136.3	& Y	&	$-$	&	$-$	\\
13 & 61.076	&	56.843	&	3.60$\pm$0.58		&		133.4	& Y	&	$-$	&	$-$	\\
14 & 61.050 &	56.933	&	4.92$\pm$1.58		&		137.3	& Y	&	4.6$\pm$0.3	&	136	\\
15 & 61.050 &	56.960 	&	4.06$\pm$1.32		&		134.5	& Y	&	4.0$\pm$0.7		&	134	\\
16 & 61.046	&	56.907	&	4.36$\pm$0.96		&		136.4	& Y	&	$-$	&	$-$	\\
17 & 61.033	&	56.972	&	3.33$\pm$0.94		&		134.7	& Y	&	$-$	&	$-$	\\
18 & 61.012	&	56.997	&	3.82$\pm$0.85		&		132.9	& Y	&	$-$	&	$-$	\\
19 & 61.007	&	56.947	&	3.33$\pm$1.02		&		132.9	& Y	&	$-$	&	$-$	\\
20	& 61.097	& 56.998 & 4.09$\pm$1.93 & 135.8 & N & $-$	&	$-$	\\
21 & 61.113 &57.001 & 3.51$\pm$1.59 & 143.5  & N & $-$	&	$-$	\\
22 & 61.178 & 56.839 & 3.58$\pm$1.85 & 140 & N & $-$	&	$-$	\\
23 &61.229 & 56.878 & 4.29$\pm$1.92 & 136.2 & N &  $-$	&	$-$	\\
24 & 61.314 & 56.870 &4.53$\pm$1.64 & 144.8 & N & $-$	&	$-$	\\
 \hline
\end{tabular}
\end{center}
\end{table*}


\section{Geometry of magnetic field}

\subsection{ Optical polarization}
We have estimated the values of linear polarization for 24 field stars, out of which we have considered 19 stars where $p/ep\geq$ 3 ($ep$ is the error in polarization values). We have presented the polarization values and the position angle of polarization of these 19 field stars of CB17 along with their right ascensions and declinations in Table-3. We have observed that the coordinates of 5 stars reported by Soam et al. (2016) are matching well with our detected stars. $p$, and $\theta$ values for \# 8, 9, 10, 14 and 15 reported by them are also in good agreement with our estimation. Columns 7 and 8 represent the polarization and position angle of polarization obtained by Soam et al. (2016), for those stars common in both the studies. As discussed earlier, they covered mainly the northern part of the cloud, whereas we have covered the south and west region of the cloud.


\begin{figure}
\begin{center}
\hspace{1cm}
\includegraphics[width=120mm]{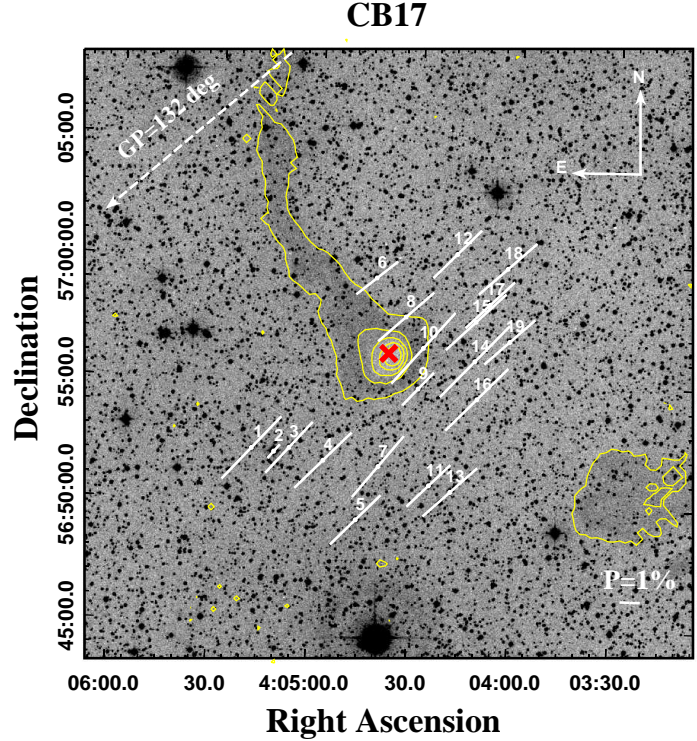}
\caption{The optical polarization vectors (blue solid lines) of 19 field stars ($p/ep>3$) obtained from our work are plotted on a 25$^\prime\times25^\prime$ DSS image of the Bok Globule CB17. A reference vector is shown at the right bottom with a polarization value of 1$\%$. The black dashed line represents the position angle of the galactic plane ($\theta_{GP}=132^\circ$). The central position of the globule (04:04:38, +56:56:12) is shown by the red cross mark. Here, we  have also plotted the contours of Herschel SPIRE 500 $\mu m$ dust continuum emissions over the DSS image, ranging from from 6 to 80 mJy beam$^{-1}$, increasing in a step size of 18 mJy beam$^{-1}$. The contour shows a cometary shape of the cloud CB17. In the contour map, a dark nebula LDN 1388
(RA = 04:03:15, DEC = +56:50:27) is located towards the south west region of CB17.}
\end{center}
\end{figure}


The mean value of polarization of these 19 stars is found to be $<p>$ = 3.52\% with a standard deviation of $\sigma_p = 0.89\%$ and the mean value of position angle is $<\theta> = 136^\circ$ with a standard deviation $\sigma_\theta = 2.84^\circ$. The results obtained by Soam et al. (2016) for 120 field stars (mainly in northern region of CB17)  were $<p> = (3.3 \pm 0.9)\%$ and $<\theta> = (137 \pm 6)^\circ$.

We have created the polarization map by plotting the polarization vectors over the DSS (Digital Sky Survey) image of CB17, as shown in Fig-1. The length of the vector shows the extent of polarization and it's orientation represents the projection of plane of the sky component of the magnetic field over that region. It is clear from Fig-1 that all the polarization vectors are aligned almost along the same direction thereby giving the alignment of the plane-of-sky component of the magnetic field. Moreover, the magnetic field orientation at the envelope ($<\theta^{env}_B> = 136^\circ$) is found to be aligned along the galactic plane over that region ($\theta_{GP}=132^\circ$ at galactic longitude $b=3.39^\circ$). Hence, there is a probability that the local magnetic field of the cloud is dominated by the Galactic magnetic field. Moreover, star \#10 of Table-3, the closest star to the center of the globule in the western region is having a limiting distance of $2.2\times 10^{4}$ AU, gives the inner scale of the envelope magnetic field. Thus the effect of Galactic magnetic field upon the envelope magnetic field of the cloud will be dominant to this distance.

The orientation of the magnetic field in the plane-of-the-sky (POS) can be estimated via the effects of aligned dust grains. The optical polarization vectors are aligned parallel to POS component of the magnetic field which is beleived to be due to the effects of dust grain alignment based on Davis \& Greenstein (1951) paramagnetic relaxation theory. But, it is known from observations that the grain alignment changes with environments and sometimes fails (Hoang \& Lazarian 2014). The actual mechanism by which dust grains align with the magnetic field has been a matter of debate for many years (Jones \& Spitzer 1967; Purcell 1979; Lazarian 2003, 2007; Andersson et al. 2015). However, the radiative torques (RATs) mechanism, initially proposed by Dolginov \& Mitrofanov (1976), is emerging out to be successful in explaining dust grain alignment in various environments (e.g., Hoang \& Lazarian, 2014; Hoang et al., 2015; Andersson et al., 2015 etc.).

In this study, we have found a very good alignment in the orientation of galactic magnetic field ($\theta_{GP}= 132^\circ$) with that of the local magnetic field ($<\theta^{env}_B> = 136^\circ$), with an offset of 4$^\circ$. Since the galactic magnetic field is dominant over the outer less dense region of the cloud, hence we cannot infer much information about the local magnetic field of the cloud from the optical polarimetric study. Thus, we have opted for sub-millimeter polarimetry.


\begin{table*}
\caption{Sub-millimetre polarization data of CB17 with $I>0$,~ $p/ep>2$ ~\&~ $ep<6.5\%$. To compare with the optical polarization angles, the submm polarization angles in column-5 have been rotated by 90 degrees to indicate the orientation of the magnetic field as shown in column-6.}
\begin{center}
\begin{tabular}{|c|c|c|c|c|c|}
\hline
   \# 			& RA(2000)       &  DEC(2000)     			&  p$^{sub}$    		&     PA     		& $\theta^{core}_B$ (=PA+90$^\circ$)  \\
						& (h m s)			  &($^\circ$ $'$ $''$) 	&	(\%) 							&	($^\circ$)					& ($^\circ$) \\
\hline
   1		     &	04:04:38.62   &  56:56:21.96   			&  21.4$\pm$2.8   &  --46.4$\pm$8.1    &	43.6 \\
	2		     &	04:04:37.98	  &	56:56:12.31   			&  17.8$\pm$6.0   &  --28.7$\pm$9.3    &	61.3 \\
	3	     	&	04:04:37.33	  &	56:56:02.65  			&  15.5$\pm$3.5   &  --10.7$\pm$5.0    &	79.3 \\
	4 		     &	04:04:37.33	  &  56:56:12.31   			&  25.8$\pm$5.4   &  --56.8$\pm$5.0    &	33.2 \\
	5		     &	04:04:37.33	  &  56:56:21.96   			&  14.1$\pm$6.1   &  --69.1$\pm$11.3    &	20.9 \\
	6		     &	04:04:36.05   &  56:56:12.31   			&  23.8$\pm$6.5   &  --63.8$\pm$7.0    &	26.2 \\
 \hline

\end{tabular}
\end{center}
\end{table*}



\begin{figure*}
\begin{center}
\vspace{.2cm}
\includegraphics[width=80mm]{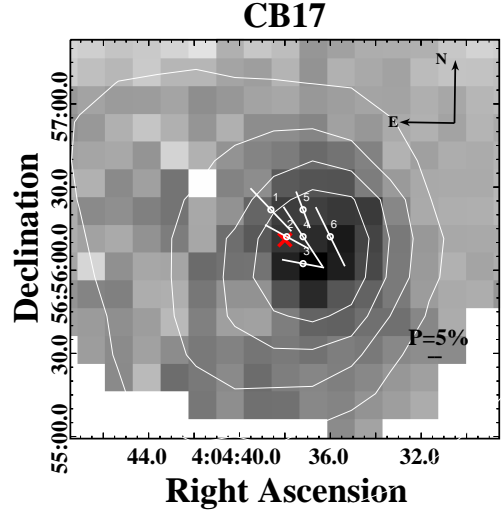}
 \vspace{.2cm}
\caption{850 $\mu m$ polarization vectors (white solid lines) are sampled on a 10$^{''}$ grid. The sub-mm polarization angles are rotated by 90 degrees to show the orientation of the magnetic field, for comparison with the optical polarization angles. The intensity data are taken from SCUPOL data, rather than SCUBA Legacy Catalog. Vectors are plotted where $I>0$, $p/ep>2$ and $ep<6.5\%$. A reference vector is shown at the right bottom with a polarization value of 5$\%$. The red cross mark shows the center of the cloud. Here, we  have also plotted the contours of Herschel SPIRE 500 $\mu m$ dust continuum emissions over the DSS image, ranging from from 6 to 80 mJy beam$^{-1}$, increasing in a step size of 18 mJy beam$^{-1}$.}
\end{center}
\end{figure*}

\subsection{Sub-millimeter Polarization}
We have taken the sub-millimeter polarization data from Matthews et al. (2009) legacy data set. The observations were performed at  James Clerk Maxwell Telescope (JCMT), Mauna Kea, Hawaii. The intensity data are taken from SCUPOL data, rather than the SCUBA Legacy Catalog. Mathews et al. (2009) limited their data to those measurements for which $I>0$, $p/ep>$2 and $ep<4\%$. In this paper, we have extended our data to the measurements for which $I>0$, $p/ep>2$ and $ep<6.5\%$. In this way, we have obtained six data points and are presented in Table-4. It does not matter to consider the value of $ep$ to be 6.5 or 7\% as both of them show the same number of data points. However, we did not consider $ep$ more than 7\% as it may include noise in the data set.
We find that two of six data points are the exact match as obtained by  Matthews et al. (2009) (\#1 and \#3 of Table-4).  To compare with the optical polarization angles, the sub-mm polarization angles (PA) have to be rotated by 90 degrees to show the orientation of the magnetic field ($\theta^{core}_B$) (Wolf et al. 2003). The mean values of polarization and polarization position angle along with their standard deviations are found to be $<p^{sub}>=19.7 \%$, $\sigma_{p} = 4.3 \%$ and $<PA> = - 45.9^\circ$ , $\sigma_{\theta}=20.5^\circ$, where as the mean magnetic field at the core is given by $<\theta^{core}_B> = 44.1^\circ$. The polarization map is presented in Fig-2, where 850 $\mu$m polarization vectors (rotated by 90 degrees) are sampled on a 10$^{''}$ grid. From the polarization map, it is clear that all the polarization vectors in the submillimeter range are almost oriented in a uniform direction over the central core of the globule, and the direction of orientation is perpendicular to that of Galactic magnetic field ($\theta_{GP}$=$132^\circ$). Moreover, the farthest distance from the center of the globule to the sub-mm polarization, \#6 is at a distance of $4\times10^3$ AU. Thus, to this range the local magnetic field of the cloud is dominant.

\section{ Magnetic Field Strength }

We have estimated the magnetic field strength for SCUPOL data using the equation (Chandrasekhar and Fermi 1953) given by,

\begin{equation}
~~~~~~~~~~~~~~~~~~~~~ B = |B_{POS}| = \sqrt{\frac{4\pi}{3}~\rho_{gas}}~\frac{v_{turb}}{\sigma_\theta},~~~~~
\end{equation}
where, $\rho$$_{gas}$ (g cm$^{-3}$) is the gas density of the cloud, \textit{v}$_{turb}$ is the rms turbulence velocity and $\sigma_\theta$ is the standard deviation of the polarization position angles in radians. Further, $\rho_{gas} = 1.36n_{H_2}M_{H_2}$ (Henning et al. 1997), where $M_{H_2}$ = 2.0158 amu = 2.0158$\times1.66\times10^{-24}$g, is the mass of H$_2$ molecule, and $n_{H_2}$, the mean volume density of the core and is taken to be 3.9$\times10^5cm^{-3}$ (Launhardt et al. 1997). Kane \& Clemens (1997) reported the characteristic turbulent velocity to be 0.13 km s$^{-1}$ for CB17, where as the characteristic thermal velocity is 0.09 km s$^{-1}$. Using equation(1), the magnetic field strength is estimated to be $\approx 99\mu$G. Uncertainty in $B$ is not estimated here because the uncertainty in the turbulent velocity is not known. However, Soam et al. (2016) obtained the value of magnetic field at the core of CB17 to be 149$\mu$G, considering only two submillimeter polarization data, whereas we have taken six polarization data in our work.

\section{Estimation of Visual Extinction $(A_V)$}

\subsection{Near-infrared Photometric Technique}
In Near-infrared (NIR) Photometric Technique, a set of dereddened colors of each star is generated from their observed reddened colors by using trial values of visual extinction ($A_V$) and the reddening law of Rieke \& Lebofsky (1985). The estimated dereddened color indices are then compared with intrinsic color indices of normal main-sequence stars to obtain the spectral type and extinction of the star. The absolute magnitude ($M_K$) corresponding to that spectral type is known from the standard literature (Cox 2000). It is to be noted that distance is not considered in that technique when calculating extinction. Actually, knowing the absolute magnitude and extinction of a star, distance is indirectly calculated. In NIR technique, three unknown quantities [extinction ($A_V$), absolute magnitude ($M_{K_S}$) and distance ($d$)] are estimated which uses the 2MASS catalogue, intrinsic colors of normal main-sequence stars and the reddening law of Rieke \& Lebofsky (1985). This technique was developed by Maheswar et al. (2010). Researchers used the NIR technique to estimate the distance of a star and finally distance of a Bok globule from the distance versus extinction plot.  Recently, Das et al. (2015) determined the distances of six small Bok globules (CB 17, CB 24, CB 188, CB 224, CB 230 and CB 240) using this technique. But this technique could not estimate distance to a star accurately due to consideration of three unknown quantities in the model.

 However, the release of \textit{Gaia} data gave the opportunity for the first time to obtain the distance of faint stars in the vicinity of a molecular cloud which may be used to estimate extinction and absolute magnitude of stars (at three filters $J$, $H$, and $K_{S}$) accurately enough using the NIR technique. The NIR data of CB17 at three different wavelengths $J(1.235\mu m)$, $H(1.662\mu m)$ and $K_S(2.159\mu m)$ of the field stars have been collected from $2MASS$ Point Source Catalog ($25'\times25'$) (Cutri et al. 2003). The absolute magnitude and extinction of a star can be calculated using this technique by knowing apparent magnitude and distance of a star. The technique developed here is now discussed below in details.

Distance modulus of a star can be written as:
\begin{equation}
~~~~~~~~~~~~X - M_X = 5 (log~d -1) + A_X
\end{equation}
where, $X$ (= $J, H$ or $K_S$), $M_X$ (= $M_J, M_H$ or $M_{K_S}$) and $A_X$ (= $A_J, A_H$ or $A_{K_S}$) are the apparent magnitude, absolute magnitude and extinction of stars at $J, H$ and $K_S$ filters respectively.

Equation (2) can be written in the form:
\begin{equation}
~~~~~~~~~~~~A_V = \frac{1}{(A_X/A_V)} \left[(X - M_X) + 5(1 - log~d)\right]
\end{equation}
It is observed that visual extinction ($A_V$) is related to $A_J, A_H$ and $A_{K_S}$. According to Rieke \& Lebofsky's (1985) extinction law,  $A_J/A_V = 0.282$, $A_H/A_V = 0.175$, $A_{K_S}/A_V = 0.112$. This extinction law was derived towards the Galactic center in the Arizona-Johnson photometric system, not 2MASS. Cambr\'{e}sy et al. (2002) later found that the slope of the reddening vector measured in the 2MASS colour-colour diagram is in good agreement with Rieke \& Lebofsky's extinction law. This set of values was also used by Maheswar et al. (2010) and Das et al. (2015) to estimate the extinction of stars using the $JHK_S$ near-infrared photometry.

Since $M_J, M_H$, and $M_{K_S}$ values are unknown quantities, $A_V$  is estimated using an iteration technique where a particular set of ($M_J, M_H$, $M_{K_S}$) values will give the best fit to equations obtained for three filters from  equation (3). In the next section, the visual extinction ($A_V$) of a star is estimated.

\subsection {Calculation of $A_V$}
The Near Infrared data have been collected from $2MASS$ Point Source Catalog (FoV: $20'\times20'$) (Cutri et al. 2003).
The data points satisfying the following conditions have been considered for the calculations.
\begin{enumerate}
	\item [(a)] Sources having quality flag (Qflag) `AAA' in all the three filters ($J, H$ \& $K$), which represents a high signal to noise ratio ($SNR>10$).
	\item [(b)] Sources with photometric uncertainties less than or equal to 0.045 in all three filters.
	\item [(c)] To avoid M-type stars, $J-K_{s} \le 0.75$ is considered. Since unreddened M-type stars situated
across the reddening vectors of A0-K7 dwarfs make it difficult to identify the reddened A0-K7 dwarfs from the
unreddened M-type normal stars (Maheswar et al. 2010).

\end{enumerate}

In this paper, we have incorporated \textit{Gaia} database (Gaia collaboration 2016, 2018) which provides the parallax measurement of stars along with their standard deviation. The data for 37 field stars of CB 17 with their \textit{2MASS} identification number is now shown in the second column of Table-5. The distance of individual star (in parsec) associated with cloud CB17 has been estimated from the parallax, and is presented in the third column along with their uncertainty.

In Fig. 3, the $(J - H)$ vs. $(H - K_S)$ color color diagram is shown for selected field stars in the vicinity of the Bok globule CB 17. The intrinsic colors of main sequence stars are taken from Maheswar et al. (2010) and Strai\~{z}ys \& Lazauskaite (2009), and red giants from Strai\~{z}ys \& Lazauskaite (2009).


\begin{figure*}
\begin{center}
\hspace{-3cm}

\includegraphics[width=120mm]{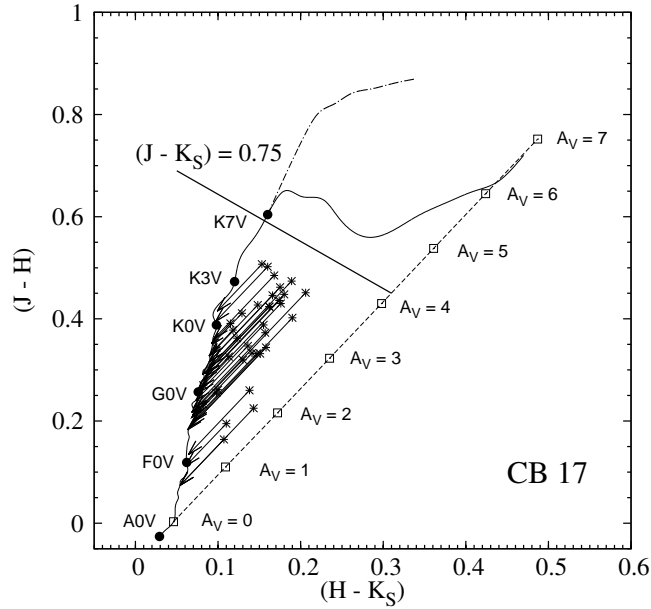}

\caption{The $(J - H)$ vs. $(H - K_S)$ color color diagram drawn for selected 37 field stars in the vicinity of the Bok globule CB17. The solid curve shows the locations of unreddened main sequence stars and dot-dashed curve represents the red giants. A dashed line (the reddening vector for an A0V type star showing visual extinction ($A_V$) from 0 to 7) is drawn parallel to the Rieke \& Lebofsky's (1985) interstellar reddening vector. The black filled circle represents the locations of the main-sequence stars. The solid line represents the upper limit $(J - K_S) \le 0.75$ set to remove M-type stars. The asterisk symbols represent the observed reddened colors  and the arrows are drawn from the observed to the final dereddened colors obtained by the iteration technique.}
\end{center}
\end{figure*}

To estimate $A_V$ of a star for a given value of reddened magnitudes ($J$, $H$ and $K_S$) and distance ($d$), one need to search for a unique set of $M_J$, $M_H$ and $M_{K_S}$ using an iteration technique which will give best fit to equations obtained for three filters from  equation (3). Then, estimated color indices ($M_H  - M_{K_S}$) and ($M_J - M_H$) obtained from this iteration are compared with intrinsic colors of main sequence stars (see Fig. 3) to obtain the spectra of the star. Computations are executed considering trial values of $M_J$, $M_H$ and $M_{K_S}$ in the range $- 10$ to $+ 20$ with a step size of 0.001. Now $A_V$ values obtained from these three equations are compared with each other. The condition adopted here is $|A_V$ (for J-filter) $- A_V$ (for H-filter)$|$, $|A_V$ (for H-filter) $- A_V$ (for K-filter)$|$, $|A_V$ (for K-filter) $- A_V$ (for J-filter)$| \leq 0.001$.  It is found from this analysis that a reddening vector is obtained with different sets of ($M_J$, $M_H$, $M_{K_S}$) and $A_V$, parallel to the Rieke \& Lebofsky's (1985) interstellar reddening vector (see Fig-3). To determine the unique set, ($M_J - M_H$) and ($M_H - M_{K_S}$) are calculated for all such sets. Then these values are compared with intrinsic colors  $(J - H)$ and $(H - K_S)$ of main-sequence stars and the best fit is obtained from this comparison. The results obtained from this study for 37 stars are shown in fourth, fifth, sixth and seventh columns of Table-5. In Fig. 3, the arrows are drawn from the observed to the final dereddened colors obtained by the iteration technique. One can calculate the uncertainty in $A_V$ from Maheswar et al. (2010).

The uncertainty in absolute magnitude $M_X$ ($M_J$, $M_H$ or $M_{K_S}$) can be calculated using the equation,
\begin{equation}
\delta M_X = \sqrt{(\delta_X)^2 + \left(2.171 \times \frac{\delta d}{d}\right)^2 + (\zeta_X \times \delta A_V)^2}
\end{equation}
where, $\delta_X$ (= $\delta_J, \delta_H$ or $\delta_{K_S}$) is the uncertainty in apparent magnitude, $\delta d$ is the uncertainty in distance and $\zeta_X$ is a constant ($\zeta_J$ = 0.282, $\zeta_H$ = 0.175 and $\zeta_{K_S}$ = 0.112). The uncertainties obtained from this work are shown in Table-5.


\begin{table*}
\caption{Estimated values of extinction ($A_{V}$) for the field stars of CB17 are presented in this table. Column-2 gives the 2MASS co-ordinate of the filed stars in ascending order. Column-3 gives the distance (in parsec) of the field stars taken from $Gaia$ database. Columns 4, 5 and 6 give the values of absolute magnitudes $M_{J}$, $M_{H}$ and $M_{K_S}$ of the field stars in J, H and $K_S$ band along with their standard deviation and column-7 gives the estimated values of $A_{V}$. The last column represents four different fields Field-1, Filed-2, Field-3 and Field-4 of CB17 denoted by F1, F2, F3 and F4 respectively.}
\begin{center}
\begin{tabular}{|c|c|c|c|c|c|c|c|}
\hline
\hline
 SN  & 2MASS & D (pc)  &	$M_{J}$	&	$M_{H}$  & $M_{K_S}$ &	$A_{V}$ (in mag) &	Field     \\
		
\hline
\hline
1	&	04032932+5651265	&	1016	$\pm$	36	&	2.672	$\pm$	0.155	&	2.277	$\pm$	0.117	&	2.180	$\pm$	0.096	&	1.00	&	F4	\\
2	&	04033156+5653440	&	914	$\pm$	17	&	2.049	$\pm$	0.146	&	1.835	$\pm$	0.100	&	1.761	$\pm$	0.072	&	0.38	&	F4	\\
3	&	04033686+5648196	&	1517	$\pm$	70	&	1.696	$\pm$	0.177	&	1.431	$\pm$	0.138	&	1.352	$\pm$	0.117	&	1.54	&	F4	\\
4	&	04034054+5654511	&	1008	$\pm$	51	&	3.021	$\pm$	0.183	&	2.690	$\pm$	0.145	&	2.599	$\pm$	0.127	&	0.11	&	F4	\\
5	&	04035141+5655522	&	969	$\pm$	19	&	1.497	$\pm$	0.152	&	1.423	$\pm$	0.104	&	1.369	$\pm$	0.075	&	1.41	&	F4	\\
6	&	04035778+5701179	&	1401	$\pm$	70	&	2.450	$\pm$	0.196	&	2.252	$\pm$	0.151	&	2.182	$\pm$	0.130	&	1.91	&	F2	\\
7	&	04035868+5649234	&	1189	$\pm$	57	&	2.870	$\pm$	0.185	&	2.582	$\pm$	0.144	&	2.496	$\pm$	0.123	&	1.49	&	F4	\\
8	&	04035893+5700276	&	878	$\pm$	16	&	2.152	$\pm$	0.140	&	1.929	$\pm$	0.096	&	1.856	$\pm$	0.069	&	0.91	&	F2	\\
9	&	04040916+5701533	&	215	$\pm$	26	&	7.136	$\pm$	0.333	&	6.913	$\pm$	0.293	&	6.845	$\pm$	0.276	&	0.30	&	F2	\\
10	&	04041127+5654284	&	252	$\pm$	37	&	5.413	$\pm$	0.347	&	5.197	$\pm$	0.330	&	5.125	$\pm$	0.322	&	0.16	&	F4	\\
11	&	04041204+5650112	&	1216	$\pm$	63	&	3.197	$\pm$	0.198	&	3.008	$\pm$	0.154	&	2.941	$\pm$	0.133	&	1.34	&	F4	\\
12	&	04041335+5653068	&	901	$\pm$	26	&	2.870	$\pm$	0.156	&	2.543	$\pm$	0.112	&	2.454	$\pm$	0.087	&	0.94	&	F4	\\
13	&	04041572+5656463	&	511	$\pm$	6	&	2.900	$\pm$	0.137	&	2.675	$\pm$	0.091	&	2.605	$\pm$	0.062	&	1.38	&	F2	\\
14	&	04041817+5648344	&	1568	$\pm$	108	&	2.582	$\pm$	0.211	&	2.167	$\pm$	0.177	&	2.068	$\pm$	0.164	&	0.86	&	F4	\\
15	&	04041834+5702225	&	1659	$\pm$	137	&	1.298	$\pm$	0.210	&	1.035	$\pm$	0.193	&	0.959	$\pm$	0.186	&	0.59	&	F2	\\
16	&	04041835+5700245	&	1244	$\pm$	37	&	1.835	$\pm$	0.132	&	1.634	$\pm$	0.099	&	1.564	$\pm$	0.081	&	1.22	&	F2	\\
17	&	04042274+5654594	&	1844	$\pm$	150	&	2.291	$\pm$	0.233	&	1.941	$\pm$	0.202	&	1.848	$\pm$	0.189	&	0.57	&	F4	\\
18	&	04042783+5700030	&	1979	$\pm$	100	&	1.513	$\pm$	0.167	&	1.401	$\pm$	0.136	&	1.340	$\pm$	0.123	&	0.78	&	F2	\\
19	&	04043153+5655035	&	1207	$\pm$	81	&	3.210	$\pm$	0.202	&	2.849	$\pm$	0.171	&	2.754	$\pm$	0.158	&	1.16	&	F4	\\
20	&	04043547+5700289	&	1775	$\pm$	115	&	2.014	$\pm$	0.200	&	1.721	$\pm$	0.168	&	1.636	$\pm$	0.154	&	1.22	&	F2	\\
21	&	04043554+5702394	&	1587	$\pm$	86	&	2.205	$\pm$	0.183	&	1.948	$\pm$	0.148	&	1.870	$\pm$	0.132	&	1.22	&	F2	\\
22	&	04043647+5700307	&	984	$\pm$	40	&	3.491	$\pm$	0.183	&	3.211	$\pm$	0.136	&	3.129	$\pm$	0.114	&	1.46	&	F2	\\
23	&	04043933+5653464	&	1925	$\pm$	214	&	1.946	$\pm$	0.292	&	1.729	$\pm$	0.263	&	1.661	$\pm$	0.252	&	2.19	&	F3	\\
24	&	04044259+5659199	&	1077	$\pm$	53	&	1.378	$\pm$	0.147	&	1.171	$\pm$	0.125	&	1.103	$\pm$	0.115	&	0.51	&	F1	\\
25	&	04044471+5700059	&	711	$\pm$	21	&	4.078	$\pm$	0.151	&	3.762	$\pm$	0.108	&	3.673	$\pm$	0.087	&	1.37	&	F1	\\
26	&	04044962+5653016	&	1107	$\pm$	52	&	3.287	$\pm$	0.176	&	2.950	$\pm$	0.138	&	2.856	$\pm$	0.120	&	0.38	&	F3	\\
27	&	04045105+5701513	&	953	$\pm$	28	&	2.815	$\pm$	0.134	&	2.604	$\pm$	0.099	&	2.535	$\pm$	0.083	&	1.14	&	F1	\\
28	&	04045625+5655124	&	764	$\pm$	21	&	3.098	$\pm$	0.126	&	2.746	$\pm$	0.093	&	2.654	$\pm$	0.076	&	0.36	&	F3	\\
29	&	04045709+5658454	&	1436	$\pm$	78	&	2.015	$\pm$	0.167	&	1.698	$\pm$	0.141	&	1.608	$\pm$	0.129	&	1.21	&	F1	\\
30	&	04050081+5648573	&	1363	$\pm$	60	&	2.391	$\pm$	0.169	&	2.151	$\pm$	0.131	&	2.078	$\pm$	0.114	&	1.00	&	F3	\\
31	&	04050697+5652149	&	1596	$\pm$	71	&	1.448	$\pm$	0.152	&	1.374	$\pm$	0.123	&	1.320	$\pm$	0.109	&	0.84	&	F3	\\
32	&	04051350+5659246	&	1241	$\pm$	47	&	2.145	$\pm$	0.160	&	1.854	$\pm$	0.121	&	1.773	$\pm$	0.101	&	0.66	&	F1	\\
33	&	04051759+5705156	&	550	$\pm$	34	&	1.574	$\pm$	0.185	&	1.391	$\pm$	0.157	&	1.328	$\pm$	0.144	&	1.51	&	F1	\\
34	&	04051989+5703033	&	1581	$\pm$	82	&	1.919	$\pm$	0.189	&	1.785	$\pm$	0.150	&	1.721	$\pm$	0.131	&	1.18	&	F1	\\
35	&	04052424+5700133	&	1175	$\pm$	37	&	2.077	$\pm$	0.150	&	1.817	$\pm$	0.111	&	1.741	$\pm$	0.089	&	0.09	&	F1	\\
36	&	04052550+5702348	&	253	$\pm$	43	&	6.375	$\pm$	0.410	&	6.275	$\pm$	0.387	&	6.212	$\pm$	0.377	&	2.21	&	F1	\\
37	&	04053546+5647482	&	1757	$\pm$	152	&	1.933	$\pm$	0.239	&	1.624	$\pm$	0.211	&	1.532	$\pm$	0.198	&	1.54	&	F3	\\

\hline
\hline

\end{tabular}
\end{center}
\end{table*}

%


\begin{figure*}
\begin{center}
\includegraphics[width=120mm]{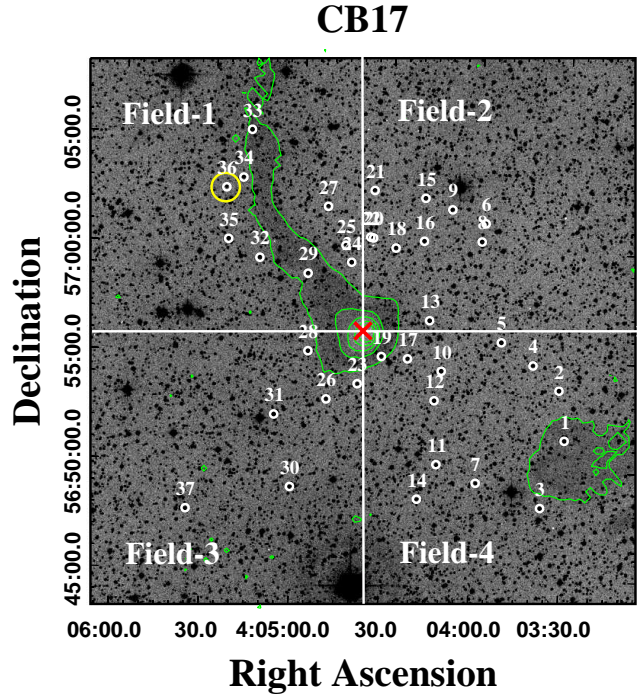}
 \vspace{.2cm}
\caption{The field of the cloud CB17 with $25'\times25'$ area is divided into four sub fields Field-1, Field-2, Field-3 and Field-4 each with area $12.5'\times12.5'$. The cross at the center represents the center of the cloud. Star \#36 (of Table-5)  is marked by a circle, which is the distance indicator of the cloud since it shows a sudden rise in extinction. We  have also plotted the contours of Herschel SPIRE 500 $\mu m$ dust continuum emissions over the DSS image, ranging from from 6 to 80 mJy beam$^{-1}$, increasing in a step size of 18 mJy beam$^{-1}$.}
\end{center}
\end{figure*}

\subsection{Distance to the cloud CB17}
A dark cloud is recognizable by the fact that, the extinction rate is higher for the stars behind the cloud than in front of it and this fact helps in estimating the distance to the clouds. The presence of interstellar cloud is detectable by a sudden increase in extinction with distance to the sources associated with the cloud (Whittet et al. 1997; Knude \& H{\o}g 1998; Alves \& Franco 2006; Lombardi et al. 2008). The method of estimating distances to the clouds is well explained in Das et al. (2015). In this method, the distance to the cloud is typically obtained from the first star
that shows a significant reddening in the \emph{Extinction} vs \emph{Distance} plot. The distance to the clouds are important to determine in order to estimate the physical properties like mass of the cloud, luminosities of the sources associated with the clouds. Several techniques have been used by researchers to estimate the distance to the clouds.


\begin{figure*}
\begin{center}
\includegraphics[width=120mm]{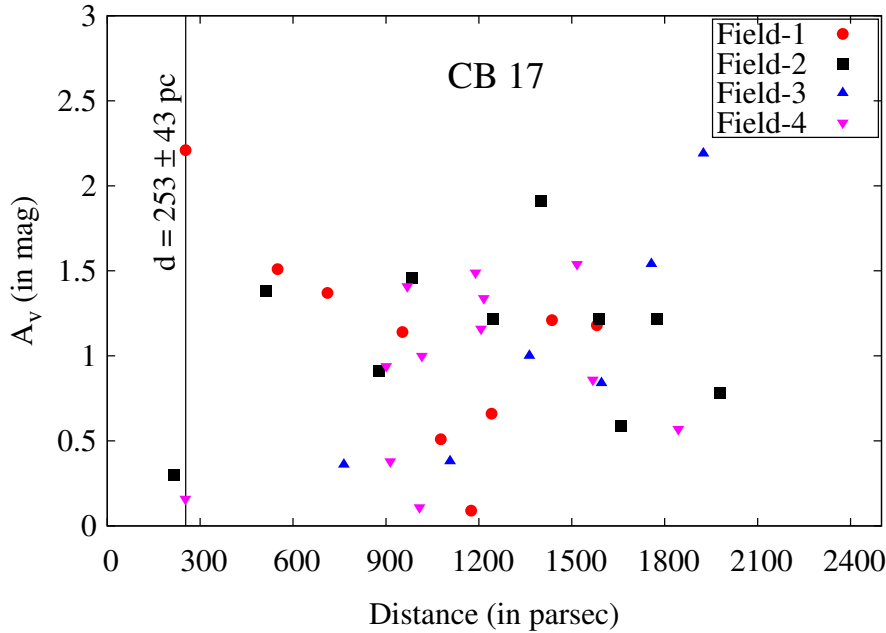}
 \vspace{.2cm}
\caption{The visual extinction ($A_V$) (in mag) is plotted against the distance, $D$ (in parsec) obtained from $Gaia$ data of the cloud CB17. The vertical solid line is drawn at a distance of 253 pc (Star \#36 of Table-5) where a sudden rise in $A_V$ occurs.}
\end{center}
\end{figure*}

We have divided the CB17 region into four fields Field-1, Field-2, Field-3 and Field-4 (Fig. 4)  and the stars corresponding to each field are detected which is represented in column-8 of Table-5. Then, we have generated $A_V$ vs $d$ plot (column-3 and column-7 of Table-5) which is shown in Fig. 5. In this figure, a sudden rise in extinction is observed at $d = 253 \pm 43$ pc (star \#36 corresponding to Field-1) whereas the star \#9 and \#10 having a distance of 215 and 252 parsec show low extinction values of 0.30 and 0.16 magnitude. Thus the star \#36 may be considered to be the distance indicator of the cloud as for being the first star showing significantly high extinction which is located at the edge of the cloud (please see Fig. 4). In literature, the distance of the cloud CB17 was reported to be $250 \pm 50$ pc by Launhardt et al. (2010) and 478 $\pm$ 88 pc by Das et al. (2015). This study shows that the distance obtained using the extinction-based method is in good agreement with the distance suggested by Launhardt et al. (2010) which was based on results obtained for possible associations of CB 17 with both the Lindblad ring and HD 25347. In this work, we have introduced the $Gaia$ data for the first time to estimate the distance to the cloud CB17 and have revisited the distance.

It is to be noted that the high extincted stars were not found between 253 and 511 parsec in the extinction envelope of CB17, so it is little difficult to bracket the cloud distance up to 253 parsec. Normally, the small surface area of the cloud CB17 (and other isolated Bok globules) implies that usually, only a very small number of stars intersects with the low-moderate extinction envelope (where background stars can still be detected). The supposed foreground stars \#9 and \#10 are having an offset from the line of sight of the dust emission region might not be indicative since not intersecting with the line of sight towards the extincting area. Hence the photometric method does not always provide very reliable and accurate distance estimates towards small Bok globules, like CB17.

%
\section{Relative orientation between various quantities of the cloud}
Several studies related to the correlation between magnetic field orientation, outflow direction, and minor axis were made in the past to understand the star formation processes of the clouds in a detailed manner. However, the correlation between these three parameters is still under debate. Certain studies have revealed a very good alignment between magnetic field orientation and outflow direction (Cohen et al. 1984; Vrba et al. 1986; Jones and Amini 2003; Wolf et al. 2003; Hull et al. 2013; Bertrang et al. 2014; Soam et al. 2015; Das et al. 2016), whereas misalignment between the same has also been reported in several studies (e.g., Wolf et al. 2003; M\'{e}nard and Duch\'{e}ne 2004;  Targon et al. 2011; Krumholz et al. 2013; Soam et al. 2015). Based on MHD simulations, Matsumoto and Tomisaka (2004) and Matsumoto et al. (2006) found that the alignment between the magnetic field direction and outflow direction depends on the strength of the magnetic field, i.e., stronger the magnetic field, better the alignment. Curran and Chrysostomou (2007), based on the study of 16 high-mass star forming regions, found no correlation between mean magnetic field and outflow direction, although they noticed some alignments. Targon et al. (2011), based on the results from optical polarization observations of protostars, obtained misalignment between magnetic field direction and outflow axis. Hull et al. (2013) also reported misalignment of outflows with respect to core-scale magnetic fields, while Chapman et al. (2013) found evidence for such an alignment. However, from the study of correlation of magnetic fields with bipolar outflows, Hull et al. (2014) found that the sources with low polarization fractions show indication that outflows are preferentially perpendicular to small scale magnetic fields. Krumholz et al. (2013) reported that magnetic field lines and rotational (or minor) axes are randomly aligned in cloud cores. But Chapman et al. (2013) reported a positive correlation between mean magnetic field direction and pseudodisk symmetry axis (minor axis) from the analysis of the 350$\mu m$ polarization data for the seven cores. Das et al. (2016) studied the correlation of mean magnetic field with the minor axis and outflow direction for a large globule CB34. They found a good alignment with the minor axis and mean magnetic field in one core (C1), whereas the magnetic field is perpendicular with the minor axis in other core (C2). Further, the magnetic field of core C2 is observed to be almost perpendicular with molecular outflow of CB34 and similar feature has been observed for three clouds B335, CB230 and CB68 (Wolf et al. 2003, Bertrang et al. 2014).

We attempted to find out a correlation between the mean magnetic field with the outflow and the minor axis of the cloud CB17. Relative orientations between various quantities of CB17 are presented in first row of Table-6 along with a comparative study of the same for some dark clouds. First column gives the cloud ID and column-2 to column-6 give the position angles of mean magnetic field at the envelope ($<\theta^{env}_B>$), mean magnetic field at the core ($<\theta^{core}_B>$), outflow axis ($\theta_{out}$), minor axis ($\theta_{min}$) of the core of the cloud and galactic plane ($\theta_{GP}$) respectively. $<\theta^{env}_B>$ of CB17 is found to be almost aligned along the galactic plane over that region of the sky (column-7), which indicates the dominance of galactic magnetic field over the envelope magnetic field of the cloud and thus we can not infer much about the magnetic field structure from the optical study. The similar feature has also been observed in case of CB34 (Das et al. 2016); L328, L673-7 (Soam et al. 2015); CB26 (Halder et al. (on prep.)); CB3, CB246 (Ward-Thompson et al. 2009). However, $<\theta^{core}_B>$ of CB17 (obtained by sub-mm polarimetry) traced out to be perpendicular to $<\theta^{env}_B>$ (column-8); the similar phenomenon has been observed in case of CB34-C1 (Das et al. 2016); IRAM 04191 (Soam et al. 2015) and CB54 (Wolf et al. 2003). Since, in case of CB17, $<\theta^{env}_B>$ is along the galactic plane orientation, this implies that only $<\theta^{core}_B>$ (denser region) is linked with the ongoing physical phenomena in the cloud. $<\theta^{core}_B>$ is oriented perpendicular to the $\theta_{GP}$ as well (column-9); and the similar orientation has been observed in case of CB34-C1 (Das et al. 2016); IRAM 04191 (Soam et al. 2015); CB230, CB244 (Wolf et al. 2003) as well. Moreover, $<\theta^{core}_B>$ is found to be almost aligned along the minor axis of the core of the cloud, the angular offset is nearly 5.9$^\circ$ (column-10). The alignment of $<\theta^{core}_B>$ with minor axis of the cloud fits the magnetically regulated star formation model that the magnetic field should lie along the minor axis of the cloud (Mouschovias and Morton 1991; Li 1998) and the same feature has also been observed for the clouds CB34-C1 (Das et al. 2016) and IRAM 04191 (Soam et al. 2015). The angular offset between $<\theta^{core}_B>$ and the outflow axis is found to be 80.9$^\circ$ (column-11), that is the core-scale magnetic field is oriented almost perpendicular to that of outflow direction and the similar phenomenon has also been observed in case of CB34 (Das et al. 2016); CB68 (Bertrang et al. 2014); B335, CB230, CB244 (Wolf et al. 2003) and CB3 (Ward-Thompson et al. 2009). The angular offset between $\theta_{out}$ and $\theta_{min}$ is found to be 75$^\circ$ and the same feature has been observed for CB34-C1 (Das et al. 2016).

A map is generated to visualize the relative orientations of different quantities of CB17 as described above, which is presented in Fig-6. All the angles are measured from North towards East. It is to be noted that all observations suffer from projectional effects along the line of sight, so it is necessary to study the 3D structures of the magnetic field and the object itself. But it is beyond the scope of the present work.


\begin{figure*}
\begin{center}
\includegraphics[width=120mm]{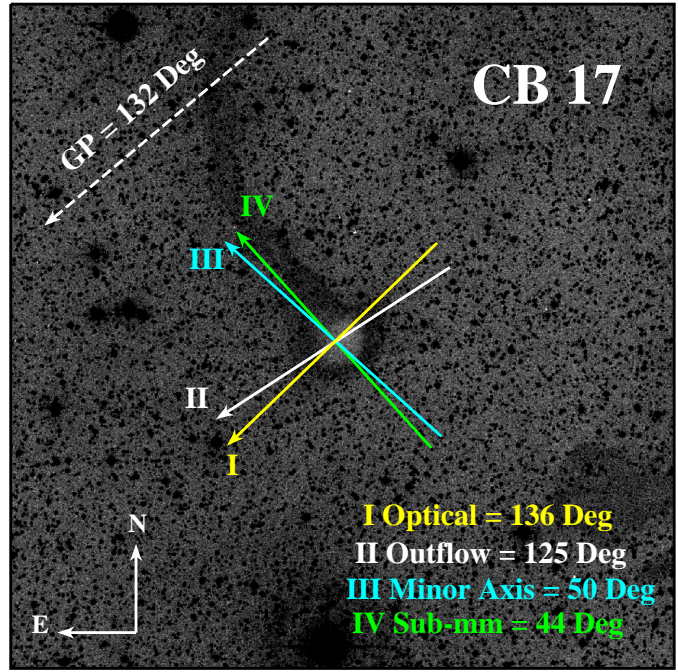}
\caption{Relative orientations of different quantities are plotted over the DSS image of CB17, where, blue(I), black(II), magenta(III) and red(IV) vectors show the position angles of envelope magnetic field (optical), core-scale magnetic field (submm), outflow axis and minor axis respectively. The black dashed line at the left corner shows the position angle of the galactic plane ($\theta_{GP}=132^\circ$) over that region of the sky.}
\end{center}
\end{figure*}

\begin{landscape}

\begin{table*}
\renewcommand{\arraystretch}{1.4}
\renewcommand{\tabcolsep}{2.4pt}

\fontsize{6pt}{8pt}\selectfont
\caption {The relative orientations between various quantities of CB17 and a comparative study of the same for some dark clouds. See the text for details.}
\begin{center}
\begin{tabular}{|p{2.2cm}|c|c|c|c|c|c|c|c|c|c|c|}
\hline
\hline
				
 Cloud & $<\theta^{env}_B>$ & $<\theta^{core}_B>$ &  $\theta_{out}$ &  $\theta_{min}$  &	$\theta_{GP}$ &	$|<\theta^{env}_B> - \theta_{GP}|$	& $|<\theta^{core}_B>-<\theta^{env}_B>|$  &	$|<\theta^{env}_B> - \theta_{GP}|$ &  	$|<\theta^{core}_B> - \theta_{min}|$ & 	 $|<\theta^{core}_B> - \theta^{out}|$ & 	 $|\theta_{out}- \theta_{min}|$  \\
	&&&&&&&&&&& \\

    & ($^\circ$) &($^\circ$) & ($^\circ$) & ($^\circ$) & ($^\circ$) & ($^\circ$) & ($^\circ$) & ($^\circ$) & ($^\circ$) & ($^\circ$) & ($^\circ$) \\

\hline
\hline

	CB17$^{a}$ &	136   & 44.1 	& 125$^\dag$   &	50$^{d}$ & 132  & 4 & 91.9 & 87.9 & 5.9	&	80.9  & 75  \\
\hline
		
	CB34-C1$^b$ &	143   & 46.7 	& --15   &	61 & 148.7 & 5.7 & 96.3 & 102 & 14.3 & 61.7  & 76  \\

CB34-C2$^b$ & 143   & 90.4 	& --15   &	--1 & 148.7 & 5.7  & 52.6 & 58.3 & 91.4 & 105.4  & 14  \\

IRAM 04191$^c$ & 112   & 44	& 28   &	30 & 139  & 27 & 68  & 95  & 14  & 16   & 2   \\

L1521F$^c$ & 22    & $-$ 	& 75    &	82  & 135 & 113 & $-$ & $-$ & $-$  & $-$ & 7   \\

L328$^c$ & 44    & $-$ 	& 20    &	$-$ & 29   & 15 & $-$ & $-$ & $-$  & $-$ & $-$   \\

L673-7$^c$ & 47   & $-$ 	& 55    &	0  & 28  & 19& $-$ & $-$ & $-$  & $-$ & 55     \\

L1014$^c$ & 15    & $-$ 	& 30    &	70  & 57 & 42  & $-$ & $-$  & $-$ & $-$  & 40   \\

L1415$^d$ & 155    & $-$ 	& $-$   &	$-$ & 123  & 32 & $-$ & $-$ & $-$ & $-$ & $-$   \\

CB68$^e$ & $-$   & 78.9  	& 142    &	$-$ & 38  & $-$ & $-$ & 40.95 & $-$  & 63.05   & $-$  \\

B335$^f$ &  $-$   & 3  	& -80    &	$-$ & 29 & $-$  & $-$   & 26 & $-$ & 83   & $-$   \\

CB230$^f$ &    $-$  & --67  	& 0    &	$-$ & 44 & $-$  &  $-$  & 111 & $-$ & 67 & $-$   \\

CB244$^f$ &   $-$   & --22  	& 45    &	$-$ & 71 & $-$  &  $-$  & 93 & $-$ & 67 & $-$   \\

CB26$^f$ & 148.3$^h$    & --65  	& --29    &	$-$ & 142 & 6.3  & 213.3  & 207 & $-$ & 36   &  $-$    \\

CB54$^f$ & 115.96$^i$    & 22  	& 30    &	$-$ &  152$^e$ & 36.04  & 93.96  & 130 & $-$ & 8   & $-$   \\

CB3$^g$ & 72    & 69  & 0$^j$ 	& 111   &	85 & 13  & 3  & 16  & 42  & 69  & 111    \\

CB246(East part)$^g$ & 60   &  $-$  	&  $-$   &	32 &  77  &  17 & $-$ & $-$ & $-$ & $-$ & $-$   \\

CB246(West part)$^g$ & 94    &   $-$	&  $-$   &	67  &  77  &  17 & $-$ & $-$ & $-$ & $-$ & $-$  \\

\hline
\hline

\end{tabular}

	
	\raggedright References. $^\dag$ Chen et al. (2012), $^{(a)}$ Our work, $^{(b)}$ Das et al. (2016), $^{(c)}$ Soam et al. (2015), $^{(d)}$ Soam et al. (2016), $^{(e)}$ Bertrang et al. (2014), $^{(f)}$ Wolf et al. (2003), $^{(g)}$ Ward-Thompson et al. (2009), $^{(h)}$ Halder et al. (on prep.), $^{(i)}$ Sen et al. (2005), $^{(j)}$ Yun \& Clemens (1994)


\end{center}

\end{table*}

\end{landscape}

\section{Conclusions}

\begin{enumerate}
    \item[(1)] We present imaging polarimetric observation of a Bok Globule CB17 in optical wavelength. The observation was carried out by 104cm Sampurnanand Telescope in R-Band at Aryabhatta Research Institute of observational sciencES (ARIES), Nainital, India. The mean value of polarization and position angle of polarization are found to be $<p>$ = 3.52\% and $<\theta> = 136^\circ$ and the corresponding standard deviations are found to be $\sigma_p = 0.89\%$ and $\sigma_\theta = 2.84^\circ$ respectively.

    The mean polarization and mean magnetic field at the core are found to be $<p^{sub}>$ = 19.7\% and $<\theta^{core}_B> = 44.1^\circ$ after reanalyzing the submillimeter data. The corresponding standard deviations are found to be $\sigma_{p}$=4.3\% and $\sigma_{\theta}=20.5^\circ$ respectively.

    \item[(2)] The direction of envelope magnetic field is found to be almost aligned with that of the galactic magnetic field of the cloud, which indicates that the envelope magnetic field is dominated by the galactic magnetic field in the less dense region of the cloud. Further, the alignment of core-scale magnetic field (traced in submm wavelength) is found to be almost 90$^\circ$ with that of the envelope magnetic field (traced in optical wavelength). Thus  only the core-scale magnetic field (denser region) is linked with the ongoing physical phenomena in the cloud. The similar phenomena has also been observed in case of CB34 (Das et al. 2016).
		
The limiting distance of the inner scale of the envelope magnetic field is traced out to be $2.2\times 10^{4}$ AU, that is to this distance the envelope magnetic field of the cloud is influenced by the galactic magnetic field, whereas, the limiting distance of the outer scale of the core scale magnetic field is traced out to be $4\times10^3$ AU. The transition between the two regimes is about $10^3-10^{4}$ AU.

    \item[(4)] The magnetic field of the core of the cloud is also found to be almost perpendicular to the outflow axis, the same phenomena has also been observed in case of B335, CB230, CB244, CB3, CB68 and CB34 (Wolf et al. 2003; Ward-Thompson et al. 2009; Bertrang et al. 2014; Das et al. 2016). The magnitude of magnetic field strength at the core of the cloud is found to be $\approx 99 \mu$G.

  \item[(5)] The direction of the core-scale magnetic field is found to be almost aligned along the direction of the minor axis of the core of the cloud, with an offset of $\approx 6^\circ$. This feature fits with the magnetically dominated star formation model and the same result has also been observed for the cloud CB34 (Das et al. 2016), and IRAM 04191 (Soam et al. 2015).

    \item[(6)] The offset between the position angle of the minor axis of the core and outflow direction is found to be 75$^\circ$. The similar orientation angle was also observed in CB34-C1 (Das et al. 2016).

\item[(7)] Using the Near Infra Red Photometric technique and \emph{Gaia} data, the distance to the globule CB 17 is obtained to be $253 \pm 43$ pc, which is in good agreement with the distance suggested by Launhardt et al. (2010). Usually, only a very small number of stars intersects with the low-moderate extinction envelope of small isolated Bok globules (like CB17), so the distance obtained from  $A_V$ vs $d$  diagram may not be always acceptable for such globules.

\end{enumerate}

\section{Acknowledgements}
We are thankful to ARIES, Nainital for providing us the Telescope time. The anonymous reviewer of this paper is highly acknowledged for his/her comments and suggestions which definitely helped to improve the quality of the paper. This work makes use of data products from the CADC repository of the SCUBA Polarimeter Legacy Catalogue and is highly acknowledged. We also acknowledge the use of the VizieR database of astronomical catalogues namely Two Micron All Sky Survey (2MASS), which is a joint project of the University of Massachusetts and the Infrared Processing and Analysis Center/California Institute of Technology, funded by the National Aeronautics and Space Administration and the National Science Foundation. \emph{Herschel} SPIRE 500$\mu m$ map of CB17 has been downloaded from the Herschel Science Archive and is highly acknowledged. This work also has made use of data from the European Space Agency (ESA) mission $\textit{Gaia}$ (\verb"https://www.cosmos.esa.int/gaia"), processed by the $\textit{Gaia}$ Data Processing and Analysis Consortium (DPAC, \verb"https://www.cosmos.esa.int/web/gaia/dpac/consortium"). Funding for the DPAC has been provided by national institutions, in particular the institutions participating in the \textit{Gaia} Multilateral Agreement. The author G. B. Choudhury wants to acknowledge Department of Science
and Technology (DST), Government of India for the DST INSPIRE fellowship (IF170830).



\end{document}